\documentclass[journal]{IEEEtran}

\ifCLASSINFOpdf
\else
   \usepackage[dvips]{graphicx}
\fi
\usepackage{url}

\usepackage{color,array,amsthm}
\usepackage{tabularx}
\usepackage{graphicx}
\usepackage{authblk}
\usepackage{amsmath,amssymb,amsfonts}
\usepackage{algorithm}
\usepackage{algpseudocode}
\usepackage{graphicx}
\usepackage{textcomp}
\usepackage{xcolor}
\usepackage{arydshln}
\usepackage{setspace}
\usepackage{lipsum}
\usepackage{textcomp}
\usepackage{indentfirst}
\usepackage{accents}
\usepackage{soul}
\setuldepth{we}
\usepackage{enumitem}
\usepackage[font=footnotesize,labelfont=bf]{caption}
\usepackage{subcaption}
\newlength{\fudgeheight}

\usepackage{hyperref,tablefootnote,footnotehyper}
\newcommand{\thickhat}[1]{\mathbf{\hat{\text{$#1$}}}}
\newcommand{\thickbar}[1]{\mathbf{\bar{\text{$#1$}}}}
\newcommand{\thickdot}[1]{\mathbf{\dot{\text{$#1$}}}}
\newcommand{\thicktilde}[1]{\mathbf{\tilde{\text{$#1$}}}}
\newcommand{\thickbreve}[1]{\mathbf{\breve{\text{$#1$}}}}

\newcommand{\bbC}{{\mathbb{C}}}
\newcolumntype{a}{>{\hsize=1.3\hsize}X}
\newcolumntype{s}{>{\hsize=0.7\hsize}X}
\newcolumntype{P}[1]{>{\centering\arraybackslash}p{#1}}
\newcolumntype{L}[1]{>{\arraybackslash}p{#1}}

\begin{document}

\title{Data-Driven Target Localization: Benchmarking Gradient Descent Using the Cram\'er-Rao Bound}

\author{Shyam Venkatasubramanian, \IEEEmembership{Student Member, IEEE}, Sandeep Gogineni, \IEEEmembership{Senior Member, IEEE}, \\ \vspace{-2ex} Bosung Kang, \IEEEmembership{Member, IEEE}, Muralidhar Rangaswamy, \IEEEmembership{Fellow, IEEE}
\thanks{This work was completed when Shyam Venkatasubramanian was an intern at AFRL, and is supported in part by the AFOSR under award FA9550-21-1-0235. Dr. Muralidhar Rangaswamy and Dr. Bosung Kang are supported by the AFOSR under project 20RYCORO51. Dr. Sandeep Gogineni is supported by the AFOSR under project 20RYCOR052.}
\thanks{Shyam Venkatasubramanian is with the Department of Electrical and Computer Engineering, Duke University, Durham, NC 27705, USA, E-mail: (sv222@duke.edu); Sandeep Gogineni is with Information Systems Laboratories Inc., Dayton, OH 45431; Bosung Kang is with University of Dayton, Dayton, OH 45469, USA (bosung.kang@udri.udayton.edu); Muralidhar Rangaswamy is with the US Air Force Research Laboratory, Wright-Patterson Air Force Base, OH 45433, USA, E-mail: (muralidhar.rangaswamy@us.af.mil).}}

\markboth{}{}
\maketitle

\begin{abstract}
In modern radar systems, precise target localization using azimuth and velocity estimation is paramount. Traditional unbiased estimation methods have utilized gradient descent algorithms to reach the theoretical limits of the Cram\'er Rao Bound (CRB) for the error of the parameter estimates. As an extension, we demonstrate on a realistic simulated example scenario that our earlier presented data-driven neural network model outperforms these traditional methods, yielding improved accuracies in target azimuth and velocity estimation. We emphasize, however, that this improvement does not imply that the neural network outperforms the CRB itself. Rather, the enhanced performance is attributed to the biased nature of the neural network approach. Our findings underscore the potential of employing deep learning methods in radar systems to achieve more accurate localization in cluttered and dynamic environments.
\end{abstract}


\IEEEpeerreviewmaketitle

\section{Introduction} \label{Sec1}

\IEEEPARstart{A}{ccurate} target localization is a critical component in contemporary radar systems, with applications that range from defense and surveillance to automotive navigation and weather prediction~\cite{blackman2004tracking, rajagopalan2008sensor, ort2020weather}. The task of estimating target parameters including azimuth and velocity with high precision is vital, calling for refined signal processing techniques. In this area, gradient descent algorithms have been a primary method, comprising iterative approaches that closely approximate the theoretical limits set by the Cram\'er Rao Bound (CRB)~\cite{Kay1993, Moore2008MLE, cheng2019CRB, xi2020MLE}. These algorithms are recognized for their robustness and efficiency in achieving the CRB.

In this study, we examine the impact of utilizing a biased optimization approach to estimate target parameters, specifically focusing on the reduction of mean squared error (MSE). This approach, despite yielding improved accuracies, should not be interpreted as exceeding the theoretical constraints of the CRB. Building upon this notion, our previous research explored the potential of regression convolutional neural networks (CNNs) in radar target localization \cite{Shyam2023datadriven}. We contrasted the performance of these networks with a classical `peak cell midpoint' method, which relied on aggregating test statistics \cite{scharf_adaptive, kraut_adaptive, vanveen_el88} into a multidimensional tensor to determine the target location using the center of the dominant cell. By extracting and synthesizing complex patterns inherent within these tensors, our proposed CNN achieved substantial gains in localization accuracy over this classical `peak cell midpoint' approach \cite{Shyam_STAP}.

Expanding upon these initial findings, we juxtapose gradient descent algorithms for target azimuth and velocity estimation with our proposed regression CNN architecture in a realistic simulated example scenario. Our analysis assesses the parameter estimation accuracies of both methodologies, underscoring the nuanced capability of the CNN model to deliver parameter estimates with a reduced MSE, attributed to its inherent bias. Through this exploration, we aim to illuminate the impact and performance improvements that data-driven models impart on the domain of radar target localization.

The structure of the paper is as follows. Section \ref{Sec2} details the representative scenario considered in this analysis. Section \ref{Sec3} provides a detailed exposition of the signal model. Section \ref{Sec4} outlines the error metric and empirical evaluations. Section \ref{Sec5} details the gradient descent algorithms utilized for parameter estimation. Section \ref{Sec6} outlines the CRB calculation. Section \ref{Sec7} presents the empirical results, and we conclude the paper in Section \ref{Sec8}, summarizing the key insights gained from our analysis and outlining potential avenues for future research.

\section{RFView\textsuperscript{\tiny\textregistered} Example Scenario} \label{Sec2}
Developed by ISL Inc, RFView\textsuperscript{\tiny\textregistered} \cite{gogineni_RFView} is a knowledge-aided, high-fidelity, site-specific and physics-based RF modeling and simulation environment that operates on a world-wide database of terrain and land cover data. Using RFView\textsuperscript{\tiny\textregistered}, we can define synthetic example scenarios that accurately model real-world environments --- our considered example scenario consists of a stationary airborne radar platform above coastal Southern California. RFView\textsuperscript{\tiny\textregistered} aggregates the information on land types, the geographical characteristics across the simulation region, and the radar parameters when simulating the radar return. The site and radar parameters of our RFView\textsuperscript{\tiny\textregistered} example scenario are given in Table \ref{radar parameters}, where we utilize a single-channel transmitter and an $L$-channel receiver. The radar return is beamformed for each size $(48/L \times 5)$ receiver sub-array, which condenses the receiver array to size $(L \times 1)$. The radar operates in `spotlight' mode and points toward the center of the simulation region. Each radar return data matrix comprises $\Lambda$ transmitted pulses. We set the elevation angle to $\phi = 0$, because in our scenario, the radar platform is far from the ground scene.

\begin{table}[h!]
\caption{Site and Radar Parameters}
\label{radar parameters}
\centering \small
\begin{tabularx}{\columnwidth}{L{0.25\textwidth-2\tabcolsep}|L{0.25\textwidth-2\tabcolsep}}
\hline Parameters & Values \\ \hline
Carrier frequency $(f_\text{c})$ & $10,000 \ \text{MHz}$ \\
Bandwidth $(B)$ \& PRF $(f_{PR})$ & $5 \ \text{MHz}$ \& $1100 \ \text{Hz}$  \\
Receiving antenna & $48 \times 5$ (horizontal $\times$ vertical) \\
Transmitting antenna & $48 \times 5$ (horizontal $\times$ vertical) \\
Antenna element spacing & $0.015 \ \mathrm{m}$ \\
Platform height & $1000 \ \mathrm{m}$ \\
Platform latitude, longitude & $32.4275^{\circ},-117.1993^{\circ}$\\
Area range $(r_{\text{lower}},r_{\text{upper}})$ & $(14538 \ \text{m},14688 \ \text{m})$ \\
Area azimuth $(\theta_{\text{min}},\theta_{\text{max}})$ & $(20^{\circ},30^{\circ})$ \\
Area velocity $(v_{\text{min}},v_{\text{max}})$ & $(175 \ \text{m/s},190 \ \text{m/s})$
\end{tabularx}
\end{table}

\begin{figure}[h!]
    \centering
    \includegraphics[width=0.95\linewidth]{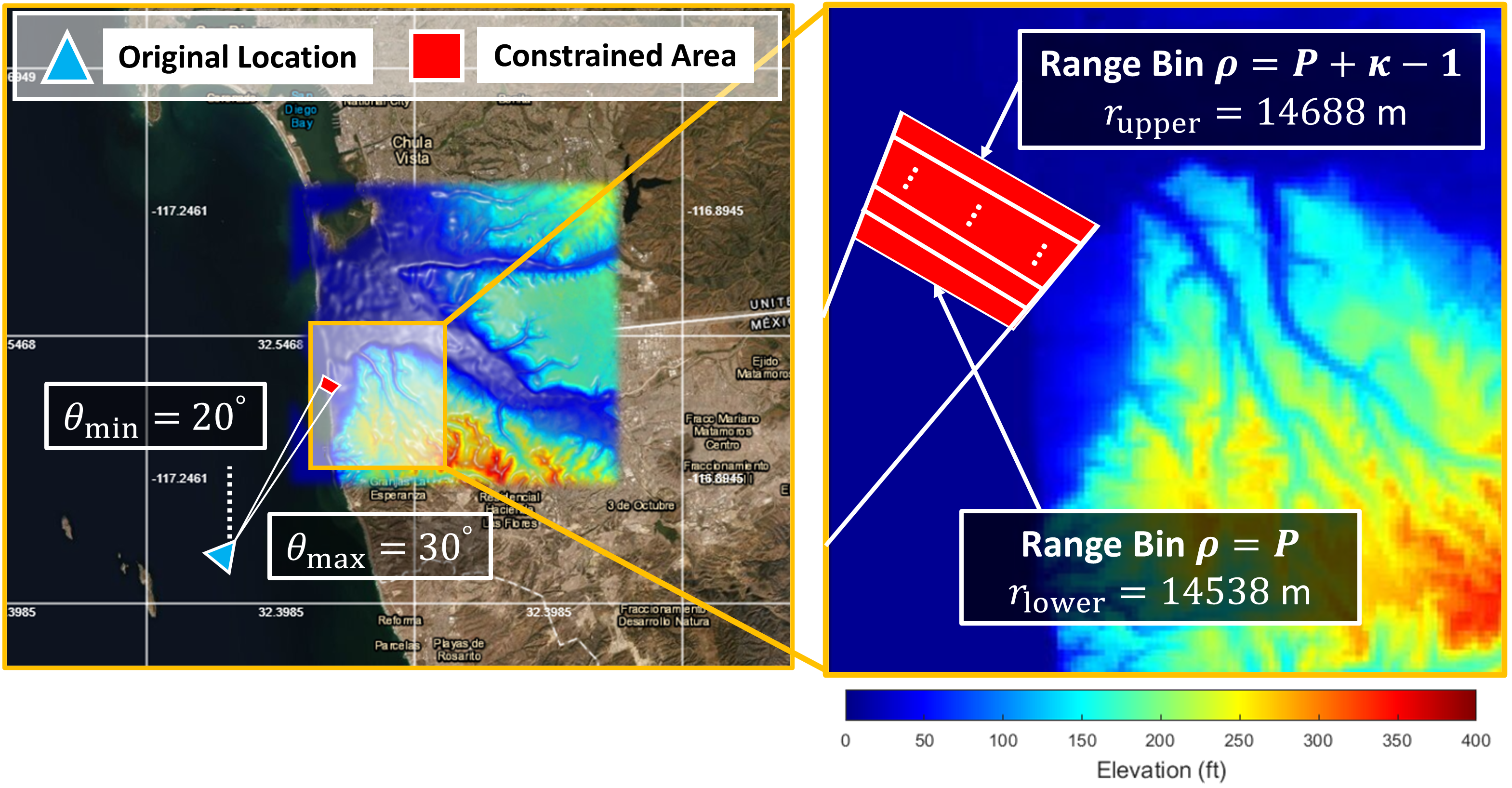}
    \caption{The RFView\textsuperscript{\tiny\textregistered} example scenario map. The blue triangle is the platform location and the red region is the radar processing area for tatget placement.}
    \label{matched map}
\end{figure}

For our RFView\textsuperscript{\tiny\textregistered} example scenario, we consider a stationary airborne radar platform within the scene described by Figure \ref{matched map}. We randomly place a moving point target in a radar processing region that contains $\kappa$ range bins and varies in range, $r$, where $r \in [r_{\text{lower}},r_{\text{upper}}]$, azimuth angle, $\theta$, where $\theta \in [\theta_{\text{min}},\theta_{\text{max}}]$, and velocity, $v$, where $v \in [v_{\text{min}},v_{\text{max}}]$. The size of each range bin is $\Delta r = \frac{c}{2B} = 30 \ [\text{m}]$, where $c$ is the speed of light and $B$ is the radar waveform bandwidth. The target RCS, $\sigma$, is randomly sampled from a uniform distribution, $\sigma \sim \mathcal{U}[\mu-l/2,\mu+l/2]$. 

We conduct a series of $N$ independent experiments, wherein each experiment involves positioning a point target uniformly at random by following the above procedure. For every target placement, we generate $K$ independent random realizations of the radar return using RFView\textsuperscript{\tiny\textregistered}. The parameters pertaining to the radar processing area are listed in Table \ref{radar parameters}.

\section{Signal Model} \label{Sec3}
We now consider a radar with an $L$-element receiver array, and $\Lambda$ transmitted pulses. Let $\mathbf{Y_\rho}\in \bbC^{(\Lambda L) \times K}$ denote a matrix comprising $K$ independent realizations of the radar return and let $\mathbf{Z_\rho}\in \bbC^{(\Lambda L) \times K}$ be a matrix consisting of $K$ independently generated clutter-plus-noise returns, both of which have been matched filtered to range bin $\rho \in \{P,P+1,...,P+\kappa-1\}$. 

Subsequently, we consider a deterministic signal, $\mathbf{S}^H_\rho \in \bbC^{K}$, in the presence of clutter, $\mathbf{c_\rho} \in \bbC^{\Lambda L}$, and noise, $\mathbf{n_\rho} \in \bbC^{\Lambda L}$. We let $\mathbf{C}_\rho, \mathbf{\bar{C}_\rho} \in \bbC^{(\Lambda \cdot L) \times K}$ denote two unique matrices consisting of $K$ independent realizations of the clutter return, $\mathbf{c_\rho}$, where $\mathbf{c_\rho}$ is dominated by $\mathbf{n_\rho}$ and has covariance matrix, $\mathbf{\Sigma_c}$. We let $\mathbf{N}_\rho, \mathbf{\bar{N}_\rho} \in \bbC^{(\Lambda \cdot L) \times K}$ denote two unique matrices consisting of $K$ independent realizations of the noise, where each noise matrix is randomly sampled from $\mathbf{n_\rho} \sim \mathcal{CN}(\mathbf{0}, \mathbf{\Sigma_n})$. Finally, we let $\mathbf{a_\rho}(\theta, \phi, v) \in \mathbb{C}^{(\Lambda L)}$ be the space-time steering vector obtained for coordinates $(\theta, \phi, v)$ in azimuth, elevation, and velocity, with array steering vector, $\boldsymbol{\xi}_\rho(\theta, \phi) \in \bbC^{L}$, and Doppler steering vector, $\boldsymbol{\psi}_\rho(v) \in \mathbb{C}^{\Lambda}$, where $f_{\text{dop}} = (2 \boldsymbol{\cdot} v \boldsymbol{\cdot} f_{\text{c}}) / \text{c}$ and $k = (2\pi \boldsymbol{\cdot} f_{\text{c}}) / \text{c}$. The sub-array phase centers are: $\mathbf{z} \in \mathbb{R}^{L \times 3}$.
\begin{gather}
    \mathbf{a_\rho}(\theta, \phi, v) =  \boldsymbol{\psi}_\rho(v) \otimes \boldsymbol{\xi}_\rho(\theta, \phi) \\
    \boldsymbol{\xi}_\rho(\theta, \phi) = e^{(i)(k)\left(\mathbf{z} \, \left[\,\text{cos}(\phi) \text{cos}(\theta), \, \text{cos}(\phi) \text{sin}(\theta), \, \text{sin}(\phi)\,\right]^H\right)} \\
    \boldsymbol{\psi}_\rho(v) = \left[ e^{-(i)(2 \pi)(\frac{f_{\text{dop}}}{f_{\text{PR}}})(0)}, \dots, e^{-(i)(2 \pi)(\frac{f_{\text{dop}}}{f_{\text{PR}}})(\Lambda - 1)} \right]^H
\end{gather}
More formally, we have that\footnote{We note that all relevant data matrices (e.g., $\mathbf{Z_\rho},\mathbf{Y_\rho}$) are preliminarily mean centered (i.e., the mean of each column in $\mathbf{Z_\rho},\mathbf{Y_\rho}$ is zero). \label{note1}}:
\begin{align}
    &\mathbf{Z_\rho} = \mathbf{\bar{C}_\rho} + \mathbf{\bar{N}_\rho} \label{eq_null} \\
    &\mathbf{Y_\rho} = \mathbf{a_\rho}(\theta, \phi, v)\mathbf{S_\rho} + \mathbf{C_\rho} + \mathbf{N_\rho} \label{eq_alternative}
\end{align}
Accordingly, the sample clutter-plus-noise covariance matrix, $\mathbf{\hat{\Sigma}_\rho}$, is calculable via Eq. (\ref{eq_covariance}) --- this matrix is used to whiten both $\mathbf{a_\rho}(\theta, \phi,v)$, and the clutter-plus-noise component of $\mathbf{Y_\rho}$.
\begin{align}
    &\mathbf{\Tilde{Y}_\rho} = \mathbf{\hat{\Sigma}_\rho}{\vphantom{\Sigma}}^{-1/2}\mathbf{Y_\rho} \\
    &\mathbf{\Tilde{a}_\rho}(\theta, \phi,v) = \mathbf{\hat{\Sigma}_\rho}{\vphantom{\Sigma}}^{-1/2} \mathbf{a_\rho}(\theta, \phi,v) \\
    &\textbf{where:} \ \ \mathbf{\hat{\Sigma}_\rho} = (\mathbf{Z_\rho} \mathbf{Z_\rho}^H) / K \label{eq_covariance}
\end{align}

\noindent Subsequently, the NAMF test statistic \cite{scharf_adaptive}, $\Gamma_\rho(\theta, \phi, v) \in \mathbb{R}^+$, for coordinates $(\theta, \phi, v)$ in range bin $\rho$ is given by:
\begin{align}
    \Gamma_\rho(\theta,\phi,v)= \frac{\| \mathbf{\Tilde{a}_\rho}(\theta, \phi, v)^H \mathbf{\Tilde{Y}_\rho}\|_2^2}{[\mathbf{\Tilde{a}_\rho}(\theta, \phi, v)^H \mathbf{\Tilde{a}_\rho}(\theta, \phi, v)] \|\text{diag}(\mathbf{\Tilde{Y}}{\vphantom{\Sigma}}^H_\rho \mathbf{\Tilde{Y}_\rho})\|_2}
\end{align}
By sweeping $\mathbf{a_\rho}(\theta, \phi,v)$ over $\theta, v$ at azimuth and velocity step size, $(\Delta \theta,\Delta v)$, with $\phi = 0$, and then recording $\Gamma_\rho(\theta,\phi,v)$ at each attribute step, we generate an azimuth-Doppler heatmap matrix. Stacking these heatmap matrices over the $\kappa$ range bins comprising the radar processing area (indexed by $\rho$) yields a 3-dimensional heatmap tensor. Per Section \ref{Sec2}, the $N$ independent experiments yield $N$ heatmap tensors in total. 

\section{Evaluations and Error Metric} \label{Sec4}
In this analysis, we consider the scenario outlined in Section \ref{Sec2} across two evaluations: \textbf{[1]} over the range of mean output SCNR, where $(\overline{\text{SCNR}}_{\text{Output}})_{\rho} \in \{-20 \ \text{dB}, -15 \ \text{dB}, ..., 20 \ \text{dB}\}$ (see \cite{shyam_radar23} for SCNR calculation), and \textbf{[2]} over the number of radar returns, where $K \in \{75, 100, ..., 300\}$. We fix the clutter-to-noise ratio (CNR) to $-20 \ \text{dB}$ ($l = 10$, suitable $\sigma$ is chosen), and we let $\Lambda = 4, L = 16$. In \textbf{[1]}, we let $K = 300$, and in \textbf{[2]}, we let $(\overline{\text{SCNR}}_{\text{Output}})_{\rho} = 20 \ \text{dB}$. Subsequently, we consider the site and radar parameters provided in Table \ref{radar parameters}, with range bins of size $\Delta r = 30 \ \text{m}$, and with azimuth and velocity step sizes of $(\Delta \theta, \Delta v) = (0.4^{\circ}, 0.5 \ \text{m/s})$, where $\kappa = 5$ and each heatmap tensor is of size $5 \times 26 \times 21$. Finally, for each value of \textbf{[1]} $(\overline{\text{SCNR}}_{\text{Output}})_{\rho}$ and \textbf{[2]} $K$, we produce $N = 1 \times 10^4$ heatmap tensors, which we partition into training ($N_{\text{train}} = 0.9N$) and test ($N_{\text{test}} = 0.1N$) datasets. We train our regression CNN on each training dataset, and evaluate the efficacy of our peak cell midpoint, gradient descent, and regression CNN methods on each test dataset. These results are provided in Section \ref{Sec7}.

Per the signal model outlined in Section \ref{Sec3}, we now define the mean squared error (MSE) metric to interpret the parameter estimation errors of our peak cell midpoint $(Err_{\text{MP}})$, gradient descent $(Err_{\text{GD}})$, and regression CNN $(Err_{\text{CNN}})$ methods. Let $(\theta_i^*, v_i^*)$ denote the ground truth target azimuth and velocity for example $i$ from our test dataset. Furthermore, let $(\thickbreve{\theta}_i, \thickbreve{v}_i)$ denote the azimuth and velocity values from the midpoint of the peak tensor cell, let $(\thickdot{\theta}_i, \thickdot{v}_i)$ denote the optimal azimuth and velocity values yielded by the gradient descent algorithm, and let $(\thicktilde{\theta}_i, \thicktilde{v}_i)$ denote the azimuth and velocity values predicted by the regression CNN, for example $i$ from our test dataset. The empirical MSE of the parameter estimates are defined as:
\begin{align}
\left(Err \right)_\theta = \frac{ \sum\limits_{i = 1}^{N_{\text{test}}} ( \theta^*_i - \thickhat{\theta}_i )^2}{N_{\text{test}}} \quad \left(Err \right)_v = \frac{\sum\limits_{i = 1}^{N_{\text{test}}} ( v^*_i - \thickhat{v}_i )^2}{N_{\text{test}}}
\end{align}
where $(\thickhat{\theta}_i, \thickhat{v}_i) = (\thickbreve{\theta}_i, \thickbreve{v}_i)$ for $(Err_{\text{MP}})$, $(\thickhat{\theta}_i, \thickhat{v}_i) = (\thickdot{\theta}_i, \thickdot{v}_i)$ for $(Err_{\text{GD}})$, and $(\thickhat{\theta}_i, \thickhat{v}_i) = (\thicktilde{\theta}_i, \thicktilde{v}_i)$ for $(Err_{\text{CNN}})$. We compare these MSE estimates with the theoretical lower bound yielded by the Cram\'er Rao Bound calculation in Section \ref{Sec6}.

\section{Gradient Descent Parameter Estimation} \label{Sec5}
We now outline the gradient descent algorithm that has been used to obtain the azimuth and velocity parameter estimates. For the azimuth case, we note that the ground truth range bin, $\rho^*$, and velocity, $v^*$, of each target are known beforehand. For the velocity case, we note that the ground truth range bin, $\rho^*$, and azimuth, $\theta^*$, of each target are known beforehand.

\subsection{Azimuth Estimation} \label{Sec5.1}
The motivation behind the \textit{Gradient Descent Azimuth Estimator} algorithm is to determine the optimal azimuth estimate, $\thickdot{\theta}$, of a given target using gradient descent. Using the matched filtered radar array data matrix, $\mathbf{\Tilde{Y}_{\rho^*}}$, and an initial estimate $\thickhat{\theta}$, corresponding to the azimuth of the peak cell midpoint from the provided heatmap tensor, the algorithm operates iteratively for a total of $T$ iterations. At each step, we obtain an estimate for the amplitude and phase of $\mathbf{S}_\rho$ via least squares. The least squares solution seeks to minimize the MSE between $\mathbf{\Tilde{Y}_{\rho^*}}$ and the estimated model, $\mathbf{a_\rho}(\theta, \phi, v^*) \mathbf{\hat{c}}$, from which the optimal array of coefficients, $\mathbf{\hat{c}}\vphantom{\Sigma}^H \in \bbC^{K}$, is defined as \cite{vantrees2002optimum}:
\begin{equation}
\mathbf{\hat{c}} = \frac{\mathbf{\Tilde{a}_{\rho^*}}(\theta, \phi, v^*)^H \mathbf{\Tilde{Y}_{\rho^*}}}{\mathbf{\Tilde{a}_{\rho^*}}(\theta, \phi, v^*)^H \mathbf{\Tilde{a}_{\rho^*}}(\theta, \phi, v^*)}
\end{equation}
Subsequently, this least squares solution is used to compute the mean squared error loss, $\mathcal{L}(\theta)$, which is defined as:
\begin{equation}
\mathcal{L}(\theta) = \frac{1}{L\, K} \sum\limits_{i=1}^{L} \sum\limits_{j=1}^{K} [(\mathbf{\Tilde{Y}_{\rho^*}})_{ij} - (\mathbf{\hat{c}} \otimes \mathbf{\Tilde{a}_{\rho^*}}(\theta, \phi, v^*))_{ij}]^2
\end{equation}
Using the gradient, $\nabla_{\theta} \mathcal{L}(\theta)$ and a preset learning rate, $\alpha$, we adjust the azimuth estimate to obtain an updated estimate of $\theta$. This procedure is summarized in Algorithm \ref{azimuth_algorithm}.

\begin{algorithm}
\caption{Gradient Descent Azimuth Estimator ($\text{GD}_\thickhat{\theta}$)}
\label{azimuth_algorithm}
\begin{algorithmic}[1]
\Procedure{$\text{GD}_\thickhat{\theta}$}{$\mathbf{\Tilde{Y}_{\rho^*}}, \thickhat{\theta}, \phi, v^*, \alpha, T$}
\State $\theta \gets \thickhat{\theta}, \ \theta_{\text{min}} \gets 20, \ \theta_{\text{max}} \gets 30$
\For{$t = 1:T$} \vspace{3px}
    \State $\mathbf{\hat{c}} \gets \frac{\mathbf{\Tilde{a}_{\rho^*}}(\theta, \phi, v^*)^H \mathbf{\Tilde{Y}_{\rho^*}}}{\mathbf{\Tilde{a}_{\rho^*}}(\theta, \phi, v^*)^H \mathbf{\Tilde{a}_{\rho^*}}(\theta, \phi, v^*)}$  \vspace{1px}
    \State \small $\mathcal{L}(\theta) \gets \frac{1}{L \, K} \sum\limits_{i=1}^{L} \sum\limits_{j=1}^{K} [(\mathbf{\Tilde{Y}_{\rho^*}})_{ij} - (\mathbf{\hat{c}} \otimes \mathbf{\Tilde{a}_{\rho^*}}(\theta, \phi, v^*))_{ij}]^2$ \vspace{-1px}
    \State \normalsize $\theta \gets \theta - \alpha \boldsymbol{\cdot} \nabla_{\theta} \mathcal{L}(\theta)$ \vspace{1px}
    \State $\theta \gets \max(\min(\theta, \theta_{\text{max}}), \theta_{\text{min}})$ \vspace{1px}
\EndFor
\State \Return $\thickdot{\theta} \gets \theta$
\EndProcedure
\end{algorithmic}
\end{algorithm}

\subsection{Velocity Estimation}
As in Section \ref{Sec5.1}, the \textit{Gradient Descent Velocity Estimator} algorithm is used to determine the optimal velocity estimate, $\thickdot{v}$, of a given target via gradient descent. Using the matched filtered radar array data matrix, $\mathbf{\Tilde{Y}_{\rho^*}}$, and an initial estimate $\thickhat{v}$, corresponding to the velocity of the peak cell midpoint from the provided heatmap tensor, the algorithm operates iteratively for a total of $T$ iterations. At each step, we obtain an estimate for the amplitude and phase of $\mathbf{S}_\rho$ via least squares. The least squares solution seeks to minimize the MSE between $\mathbf{\Tilde{Y}_{\rho^*}}$ and the estimated model, $\mathbf{a_\rho}(\theta^*, \phi, v) \mathbf{\hat{d}}$, from which the optimal array of coefficients, $\mathbf{\hat{d}}\vphantom{\Sigma}^H \in \bbC^{K}$, is defined as \cite{vantrees2002optimum}:
\begin{equation}
\mathbf{\hat{d}} = \frac{\mathbf{\Tilde{a}_{\rho^*}}(\theta^*, \phi, v)^H \mathbf{\Tilde{Y}_{\rho^*}}}{\mathbf{\Tilde{a}_{\rho^*}}(\theta^*, \phi, v)^H \mathbf{\Tilde{a}_{\rho^*}}(\theta^*, \phi, v)}
\end{equation}
Subsequently, this least squares solution is used to compute the mean squared error loss, $\mathcal{L}(v)$, which is defined as:
\begin{equation}
\mathcal{L}(v) = \frac{1}{L \, K} \sum\limits_{i=1}^{L} \sum\limits_{j=1}^{K} [(\mathbf{\Tilde{Y}_{\rho^*}})_{ij} - (\mathbf{\hat{d}} \otimes \mathbf{\Tilde{a}_{\rho^*}}(\theta^*, \phi, v))_{ij}]^2
\end{equation}
Using the gradient, $\nabla_{v} \mathcal{L}(v)$ and a preset learning rate, $\alpha$, we adjust the azimuth estimate to obtain an updated estimate of $v$. This procedure is summarized in Algorithm \ref{velocity_algorithm}.

\begin{algorithm}
\caption{Gradient Descent Velocity Estimator ($\text{GD}_\thickhat{v}$)}
\label{velocity_algorithm}
\begin{algorithmic}[1]
\Procedure{$\text{GD}_\thickhat{v}$}{$\mathbf{\Tilde{Y}_{\rho^*}}, \theta^*, \phi, \thickhat{v}, \alpha, T$}
\State $v \gets \thickhat{v}, \ v_{\text{min}} \gets 175, \ v_{\text{max}} \gets 190$
\For{$t = 1:T$} \vspace{3px}
    \State $\mathbf{\hat{d}} \gets \frac{\mathbf{\Tilde{a}_{\rho^*}}(\theta^*, \phi, v)^H \mathbf{\Tilde{Y}_{\rho^*}}}{\mathbf{\Tilde{a}_{\rho^*}}(\theta^*, \phi, v)^H \mathbf{\Tilde{a}_{\rho^*}}(\theta^*, \phi, v)}$  \vspace{1px}
    \State \small $\mathcal{L}(v) \gets \frac{1}{L \, K} \sum\limits_{i=1}^{L} \sum\limits_{j=1}^{K} [(\mathbf{\Tilde{Y}_{\rho^*}})_{ij} - (\mathbf{\hat{d}} \otimes \mathbf{\Tilde{a}_{\rho^*}}(\theta^*, \phi, v))_{ij}]^2$ \vspace{-1px}
    \State \normalsize $v \gets v - \alpha \boldsymbol{\cdot} \nabla_{v} \mathcal{L}(v)$ \vspace{1px}
    \State $v \gets \max(\min(v, v_{\text{max}}), v_{\text{min}})$ \vspace{1px}
\EndFor
\State \Return $\thickdot{v} \gets v$
\EndProcedure
\end{algorithmic}
\end{algorithm}

\section{Cram\'er-Rao Bound} \label{Sec6}
In this section, we derive the Cram\'er-Rao Bound (CRB) for the aforementioned azimuth and velocity parameter estimation errors. We compare each derived CRB with the MSE yielded by each of the approaches outlined in Section \ref{Sec4}.

\begin{figure*}[t!]
    \centering
    \begin{subfigure}{0.49\textwidth}
        \includegraphics[width=\textwidth]{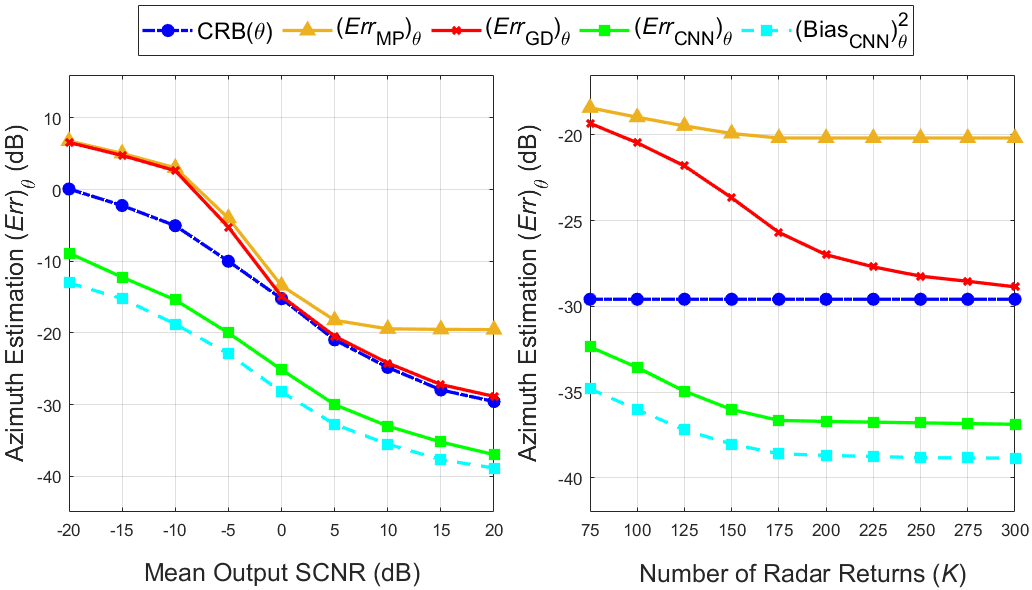}
        \caption{Azimuth Estimation Error vs. CRB --- $\left(Err \right)_\theta$ vs. $\text{CRB}(\theta)$}
        \label{fig:azim_est}
    \end{subfigure}
    \begin{subfigure}{0.49\textwidth}
        \includegraphics[width=\textwidth]{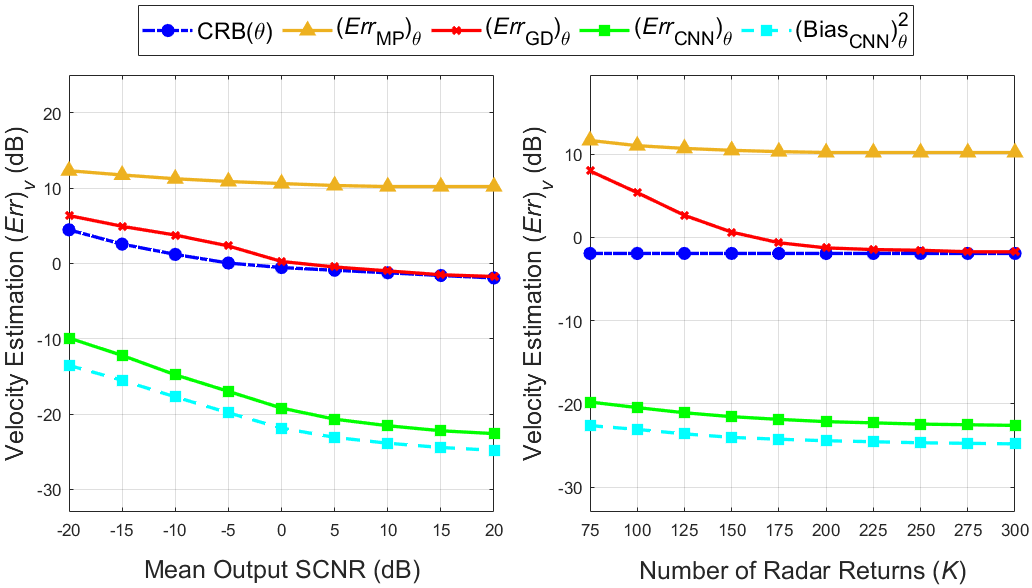}
        \caption{Velocity Estimation Error vs. CRB --- $\left(Err \right)_v$ vs. $\text{CRB}(v)$}
        \label{fig:vel_est}
    \end{subfigure}
    \caption{Comparing azimuth and velocity estimation errors of the peak cell midpoint, gradient descent, and regression CNN methods with the CRB.}
    \label{fig:est_vs_crb}
\end{figure*}

\subsection{Azimuth Estimation} \label{Sec6.1}
Deriving from the signal model in Section \ref{Sec3}, we note that $\mathbf{Y_{\rho}}$ consists of $K$ radar return realizations randomly sampled from $\mathbf{y_{\rho}} \sim \mathcal{CN}(\mathbf{b_\rho}(\theta)\thickbar{\mathbf{S}}_\rho, \mathbf{R})$, wherein $\mathbf{b_\rho}(\theta) = \mathbf{a_\rho}(\theta,\phi,v^*)$, $\thickbar{\mathbf{S}}_\rho \approx (\sum_{j=1}^K [\mathbf{S_\rho}]_j)/K$, and $\mathbf{R} \approx \mathbf{\Sigma_c} + \mathbf{\Sigma_n}$. Using the Slepian formula \cite{slepian1954estimation}, the Fisher Information for $\theta$ is defined as:
\begin{align}
    I(\theta) &= \mathbb{E}\left[ \left( \frac{\partial}{\partial \theta} \ln p(\mathbf{y_\rho} | \theta) \right)^2 \right] \\
    &= (\thickbar{\mathbf{S}}_\rho)^2 \left( \frac{\partial \mathbf{b_\rho}(\theta)}{\partial \theta} \right)^H \mathbf{R}^{-1} \left( \frac{\partial \mathbf{b_\rho}(\theta)}{\partial \theta} \right) \label{eq_slepian_azimuth}
\end{align}
Wherein $(\frac{\partial \mathbf{b_\rho}(\theta)}{\partial \theta}) = \left( \mathbf{a_\rho}(v^*) \otimes [(i)(k) \mathbf{a_\rho}(\theta,\phi) \odot [\mathbf{z} \mathbf{g}]] \right)$, with $\mathbf{g} = \left[\,-\text{cos}(\phi) \text{sin}(\theta), \, \text{cos}(\phi) \text{cos}(\theta^*), \, 0 \, \right]^H \in \mathbb{R}^3$. Therefore, the Cram\'er Rao Bound, which provides a lower bound on the MSE of any unbiased estimator of $\theta$, is defined as:
\begin{align}
    \mathbb{E}[(\hat{\theta} - \theta)^2] &\geq I(\theta)^{-1} = \text{CRB}(\theta)
\end{align}
For both of the evaluations outlined in Section \ref{Sec4}, we average $\text{CRB}(\theta)$ over the $N$ different point target locations for $\theta = \theta^*$. We compare $(Err_{\text{MP}})_\theta, (Err_{\text{GD}})_\theta,(Err_{\text{CNN}})_\theta$ with $\text{CRB}(\theta)$.

\subsection{Velocity Estimation}
Paralleling the derivation in Section \ref{Sec6.1}, we first note that $\mathbf{Y_{\rho}}$ consists of $K$ radar return realizations randomly sampled from $\mathbf{y_{\rho}} \sim \mathcal{CN}(\mathbf{b_\rho}(v)\thickbar{\mathbf{S}}_\rho, \mathbf{R})$, wherein $\mathbf{b_\rho}(v) = \mathbf{a_\rho}(\theta^*,\phi,v)$, $\thickbar{\mathbf{S}}_\rho \approx (\sum_{j=1}^K [\mathbf{S_\rho}]_j)/K$, and $\mathbf{R} \approx \mathbf{\Sigma_c} + \mathbf{\Sigma_n}$. Using the Slepian formula \cite{slepian1954estimation}, the Fisher Information for $v$ is defined as:
\begin{align}
    I(v) &= \mathbb{E}\left[ \left( \frac{\partial}{\partial v} \ln p(\mathbf{y_\rho} | v) \right)^2 \right] \\
    &= (\thickbar{\mathbf{S}}_\rho)^2 \left( \frac{\partial \mathbf{b_\rho}(v)}{\partial v} \right)^H \mathbf{R}^{-1} \left( \frac{\partial \mathbf{b_\rho}(v)}{\partial v} \right) \label{eq_slepian_velocity}
\end{align}
Wherein $\frac{\partial \mathbf{b_\rho}(v)}{\partial v} = [\mathbf{a_\rho}(v) \odot ([2 \cdot f_c] / c) \, \mathbf{h})] \otimes \mathbf{a_\rho}(\theta^*,\phi)$, with $\mathbf{h} = [(2\pi)(0)/f_{\text{PR}}, \cdots, (2\pi)(\Lambda-1)/f_{\text{PR}}]^H$. Consequently, the Cram\'er Rao Bound, which provides a lower bound on the MSE of any unbiased estimator of $v$, is defined as:
\begin{align}
    \mathbb{E}[(\hat{v} - v)^2] &\geq I(v)^{-1} = \text{CRB}(v)
\end{align}
For both of the evaluations outlined in Section \ref{Sec4}, we average $\text{CRB}(v)$ over the $N$ different point target locations for $v = v^*$. We compare $(Err_{\text{MP}})_v, (Err_{\text{GD}})_v,(Err_{\text{CNN}})_v$ with $\text{CRB}(v)$.

\section{Empirical Results} \label{Sec7}
As detailed in Section \ref{Sec4}, we compare our peak cell midpoint, gradient descent, and regression CNN methods versus the CRB across two evaluations: \textbf{[1]} over the range of mean output SCNR, and \textbf{[2]}, over the number of radar returns, $K$. For the gradient descent approach, the azimuth estimates were compiled using $\alpha = 1 \times 10^{-5}$ and $T = 100$, and the velocity estimates were compiled using $\alpha = 1 \times 10^{-2}$ and $T = 150$. For the regression CNN case, the azimuth and velocity estimates were compiled using our Doppler CNN introduced in \cite{Shyam2023datadriven}. The results of this empirical study are provided in Figure \ref{fig:est_vs_crb}.

Per Figure \ref{fig:est_vs_crb}, we observe that our regression CNN provides improved azimuth and velocity parameter estimates, yielding a test MSE that is below the CRB --- this is due to the inherently biased nature of the regression CNN. More concretely, in each of our empirical evaluations (for the regression CNN case), we observe that the bias term comprising the MSE is nonzero and large, supporting the notion that the regression CNN is a biased estimator. This evaluation underscores that the enhancements achieved with the regression CNN are not attainable utilizing conventional unbiased estimation methods.

\section{Conclusion} \label{Sec8}
In this paper, we first reviewed the performance of gradient descent algorithms for estimating azimuth and velocity in radar systems, confirming their efficiency in approaching the Cramér Rao Bound (CRB). Despite their effectiveness, we highlighted that our proposed regression CNN architecture can outperform these classical methods. In particular, our comparative analysis demonstrated that this regression CNN consistently achieves a reduced MSE when predicting target azimuth and velocity due to its biased nature; the CNN is not to be interpreted as outperforming the CRB. These findings, validated through RFView\textsuperscript{\tiny\textregistered} simulations, demonstrated the feasibility of our neural network approach in complex radar environments. Future research will be directed toward uncovering the underlying mechanisms of this improvement and optimizing neural network architectures for enhanced radar target parameter estimation.

\bibliographystyle{IEEEtran}
\bibliography{references}

\end{document}